\documentclass[12pt]{article}
 \usepackage[margin=1in]{geometry}
\usepackage{amsmath,amssymb,amsthm,graphicx}
\usepackage{url}

\usepackage{comment}
\usepackage[utf8]{inputenc}

\usepackage{makecell}
\usepackage{xcolor}
\usepackage{caption}
\usepackage{subcaption}
\usepackage[ruled,vlined]{algorithm2e}

\usepackage[shortlabels]{enumitem}
\usepackage{bbm}
\usepackage{natbib}
\setcitestyle{authoryear, open={(}, close={)}}

\newcommand{\Real}{\hbox{{I}\kern-.1667em\hbox{R}}}  
\newcommand{\vect}[1]{\mbox{\boldmath $ #1$}}

\newcommand{\Pb}{\mbox{{I}\kern-.1667em\mbox{P}}}  
\newcommand{\Ex}{\mbox{{I}\kern-.1667em\mbox{E}}}

\theoremstyle{definition}

\newcommand{\blind}{1}

\begin{document}

\def\spacingset#1{\renewcommand{\baselinestretch}%
{#1}\small\normalsize} \spacingset{1}


\if1\blind
{
  \title{\bf Time-varying Bayesian Network Meta-Analysis}
  \author{Patrick M. LeBlanc\thanks{Email: patrick.leblanc@duke.edu}\\
    Department of Statistical Sciences, Duke University\\
    and \\
    David Banks\\
    Department of Statistical Sciences,Duke University}
  \maketitle
} \fi

\if0\blind
{
  \bigskip
  \bigskip
  \bigskip
  \begin{center}
    {\LARGE\bf Title}
\end{center}
  \medskip
} \fi

\bigskip
\begin{abstract}
The presence of methicillin-resistant \textit{Staphylococus Aureus} (MRSA) in complicated skin and soft structure infections (cSSSI) is associated with greater health risks and economic costs to patients.  There is concern that MRSA is becoming resistant to other ``gold standard” treatments such as vancomycin, and there is disagreement about the relative efficacy of vancocymin compared to linezolid.  There are several review papers employing Bayesian Network Meta-Analyses (BNMAs) to investigate which treatments are best for MRSA related cSSSIs, but none address time-based design inconsistencies.  This paper proposes a time-varying BNMA (tBNMA), which models time-varying treatment effects across studies using a Gaussian Process kernel.  A dataset is compiled from nine existing MRSA cSSSI NMA review papers containing $58$ studies comparing $19$ treatments over $19$ years. tBNMA finds evidence of a non-linear trend in the treatment effect of vancomycin---it became less effective than linezolid between $2002$ and $2007$, but has since recovered statistical equivalence. 
\end{abstract}

\noindent%
\vfill

\newpage
\spacingset{2}

\section{Introduction}

\begin{figure}
    \centering
    \includegraphics[width=\textwidth]{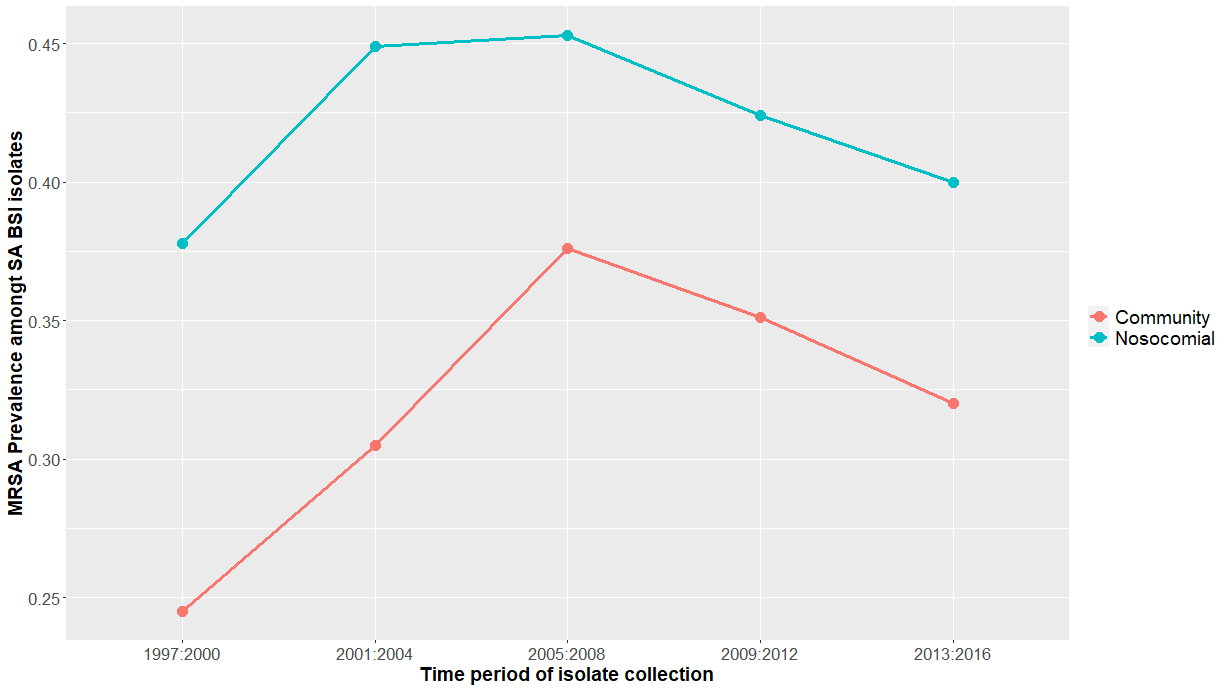}
    \caption{``SENTRY Program 20-year trends in percentage of \textit{Staphylococcus aureus} BSI isolates that are MRSA." \citep{2019_Diekema_etal} }
    \label{fig:2019_diekema_etal_mrsa_prev}
\end{figure}

Methicilin-resistant \textit{Staphylococcus aureus} (MRSA) infections are a threat to public health.  MRSA increases mortality, hospital stays, and costs \citep{2006_Crum_etal, 2007_McCollum_etal, 2012_Shorr}.  The incidence of MRSA rose globally in the late $1900$'s and early $2000$'s \citep{2008_Hersh_etal}. The SENTRY antimicrobial surveillance program, for instance, observed increasing prevalence of MRSA in complicated skin and soft structure infections (cSSSI) \citep{2007_Moet_etal}.  More recent findings suggest that MRSA prevalence peaked in $2008$ and has been declining since in the European Union and the United States \citep{2017_Klein_etal, 2019_Diekema_etal}; see Figure \ref{fig:2019_diekema_etal_mrsa_prev} for a plot of MRSA prevalence over time.  This may be because medical professionals began to implement clincial interventions to reduce the spread of MRSA \citep{2009_Liebowitz}.  Yet MRSA remains the second most common cause of antibiotic-resistant bacterial infections in the European Union \citep{2019_Gasser_Etal} and is stable in the Asia-Pacific region \citep{2019_Lim_etal}.

Growing antibiotic resistance in MRSA is a potential problem \citep{2009_Wilcox, 2009_Nathwani}.  \textit{S. aureus} is possibly developing resistance to other treatments, such as fusidic acid and mupirocin \citep{2021_Brown_etal}.  The Infectious Disease Society of America (IDSA) has long recommended vancomycin as a treatment for MRSA \citep{2012_Gould_etal}, and vancomycin is regarded as the ``gold standard" of MRSA treatments \citep{2012_Shorr}.  \citet{2007_Daum} and \citet{2004_Cosgrove_etal} state that the increase in MRSA prevalence resulted in increasing use of vancomycin and the emergence of vancomycin resistant \textit{S. aureus}. \cite{2019_Diekema_etal} finds that there was no increase in vancomycin-resistant MRSA from 2013-2016.  There remains an ``evidence gap" with respect to vancomycin-resistant \textit{S. aureus} \citep{2021_Brown_etal}.  

Many randomized controlled trials (RCT) have been conducted to assess the effectiveness of treatments for MRSA-related cSSSIs. These studies provide a mix of direct and indirect evidence for the treatments, so Bayesian network 
meta-analyses (BNMAs) have been used to estimate treatment effects; in particular, there is disagreement about whether linezolid is more effective than vancomycin \citep{2015_Thom_etal, 2016_Liu_etal, 2017_Guest_etal, 2017_McCool_etal, 2018_Li_Xu, 2019_Zhang_etal, 2019_Lan_etal, 2021_Brown_etal, 2021_Feng_etal}.  If MRSA is developing antibiotic resistance, however, treatment effects would vary across time.  The selection of treatments for a given RCT, which must take place over a short time period, will be confounded with the estimated effects of those treatments.  This type of design inconsistency \citep{2012_Higgins_etal} must be accounted for by modelling time-varying treatment effects across RCTs.   

Literature that incorporates time effects into models has focused on capturing time effects within individual RCTs rather than addressing time-based design inconsistencies.  This paper highlights some methods; \cite{2019_Tallarita_etal} provides a more exhaustive review. \cite{2011_Jansen} uses fractional polynomials to model the hazard ratio in a network of RCTs which each report the hazard ratio in some longitudinal format.  \cite{2015_Jansen_etal} generalizes this approach to other types of longitudinal data.  \cite{2016_Mawdsley_etal} proposes a framework, model-based network meta-analysis (MBNMA), which adapts methods from model-based meta analysis (MBMA) to the network setting in order to capture dose-response relationships.  \cite{2019_Pedder_etal} extends this approach to time-course models.  A common feature of these approaches is that they model time effects within individual RCTs; each RCT returns data which has a time component, e.g. dose-response curves, and the goal is to compare these time-varying functions across trials.  However, this time component does not address time-based design inconsistencies since treatment effects do not change depending on when the RCT was conducted.  

Time-based design inconsistencies could be addressed with standard meta-regression techniques \citep{2012_White_etal}. \cite{2009_Salanti_etal}, for instance, employs a meta-regression over time in a BNMA to study the effectiveness of oral health interventions: placebo treatments became more effective over time.  However, existing meta-regression BNMAs are limited to linear effects.  The true pattern of time-varying effects is unknown.  If treatments vary non-linearly, these meta-regression techniques will have limited value.  

This paper develops a class of BNMA models which can detect time-varying treatment effects: time-varying BNMA (tBNMA).  The existence of a latent, unobserved time series for treatment effects is modelled with a Gaussian Process using a combination of white noise, linear, and Matern kernels. In simulations, tBNMA outperforms existing methods when even just one treatment has time-varying effects.

The datasets of \citet{2015_Thom_etal}, \citet{2016_Liu_etal}, \cite{2017_Guest_etal}, \citet{2017_McCool_etal}, \citet{2018_Li_Xu}, \citet{2019_Zhang_etal}, \citet{2019_Lan_etal}, \citet{2021_Brown_etal} and \citet{2021_Feng_etal} are combined to form one MRSA-cSSSI dataset that includes $58$ studies comparing $19$ treatments from $2000$ to $2019$.  tBNMA detects non-linear time trends, finding that vancomycin resistance in MRSA was strongest between $2002$ and $2007$, but has since decreased.  Moreover, tBNMA finds that, while linezolid used to be significantly more effective than vancomycin, the difference is no longer statistically significant in $2019$.  

\section{Bayesian Network Meta-Analysis}

Often there are many treatment options available for a medical condition.    In a given RCT, researchers compare only a subset of those possible treatments.  To know whether a given treatment, A, is more or less effective than another treatment, B, then, there is a mix of direct evidence, where A and B are directly compared, and indirect evidence, where the treatment effect is estimated through some joint comparator C.  When there are only three treatments with two pairwise comparisons --- $A$ compared to $B$ and $B$ compared to $C$ --- then analysis is straightforward \citep{1997_Bucher_etal}. However, situations of greater complexity arise and induce a network of comparisons amongst the treatments. 

Models developed to estimate the treatment effects are referred to as Network Meta-Analyses (NMA). Frequentist NMA's have been developed in \citet{1996_Higgins_Whitehead}, \citet{2002_Lumley} and \citet{2008_Chootrakool_Shi} while Bayesian NMA's have been developed in \citet{2003_Ades}, \citet{2004_Lu_Ades} and \citet{2006_Lu_Ades}. The formulation of \cite{2011_Dias_Tech_Sup} for BNMAs with binomial data is followed in this paper. 

Let there be $I$ studies comparing (some of) $K$ treatments.  If treatment $k$ is used in study $i$, then the response variable is $y_{ik}$, the number of successes.  Each $y_{ik}$ has probability of success $p_{ik}$ for $n_{ik}$ subject.  Then $y_{ik} \, | \, p_{ik}, n_{ik} \sim \text{Bin}(p_{ik},n_{ik})$. The probabilities are modelled with a logit-link function: $\text{logit}(p_{ik}) = \mu_{i} + \delta_{i,b_i,k}1_{b_i\neq k}$.  Here, $b_i$ is the baseline treatment in study $i$.  If possible, all studies would have the same baseline, $b$, but this usually not the case, so the most common treatment is taken as the baseline.  The trial-specific effects of trial $i$ are captured by $\mu_{i}$.  These are nuisance parameters and are modelled as random effects, $\mu_i \sim N(m_\mu,\sigma_\mu^2)$.  The $\mu_i$ terms allow BNMA to  estimate the mean effect of each treatment  $d_{1k}$ even when there are unknown confounding effects between studies.

The difference in efficacy between treatment $k$ and treatment $b_i$ in study $i$ is $\delta_{i,b_i,k}$.  In a random effects model, it is drawn from a normal distribution, $\delta_{i,b_i,k} \, | \, d_{b_i,k},\sigma^2 \sim \text{N}(d_{b_i,k},\sigma^2)$. Homogeneity of variance --- that $\sigma_{b_i,k}^2 = \sigma^2$ for all $b_i$ and $k$ --- is assumed because there is not enough data to learn heterogeneous variance \citep{1996_Higgins_Whitehead}. In a multi-arm trial, the joint distribution of the $\delta_{i,b_i,k}$ is the following multivariate normal:
\[
\begin{bmatrix}
\delta_{i,b_i,2}\\
\delta_{i,b_i,3}\\
\dots\\
\delta_{ib_i,k-1}
\end{bmatrix}
\sim 
N\bigg(
\begin{bmatrix}
d_{b_i,2}\\
d_{b_i,3}\\
\dots\\
d_{b_i,k-1}
\end{bmatrix},
\begin{bmatrix}
\sigma^2 & \frac{\sigma^2}{2} & \dots & \frac{\sigma^2}{2}\\
\frac{\sigma^2}{2} & \sigma^2 & \dots & \frac{\sigma^2}{2}\\
\dots & \dots & \dots & \dots\\
\frac{\sigma^2}{2} & \frac{\sigma^2}{2} & \dots & \sigma^2
\end{bmatrix}
\bigg)
\]
It is more efficient to decompose this joint likelihood into a product of conditional likelihoods:
\[
\begin{aligned}
\delta_{i,b_i,k} \, &| \, \delta_{i,b_i,2},\dots,\delta_{i,b_i,k-1},
d_{b_i,2},\dots,d_{b_i,k-1}, \sigma^2 \\
& \sim \text{N}\bigg(d_{b_i,k} + \frac{1}{k-1}\sum_{j=1}^{k-1}[\delta_{i,b_i,j} - d_{b_i,j}],\frac{k}{(k-1)}\sigma^2\bigg)
\end{aligned}
\]

The relative difference in treatment effect between treatment $k$ and baseline $b_i$ is $d_{b_i,k}$. Under the consistency assumption \citep{2006_Lu_Ades} (also called coherence in \cite{2002_Lumley}), 
$d_{b_i,k}$ can be split into components in a way analogous to a ``differences-in-differences" approach.  The difference of $b_i$ and $k$ is equal to the the difference of the difference of $k$ and treatment $1$ (which may be taken as the general baseline $b$) and the difference of treatment $b_i$ and treatment $1$.  That is, $d_{b_i,k} = d_{1k} - d_{1b_i}$.  These baseline differences are drawn from a normal distribution: $d_{1k} \sim N(m_d,\sigma_d^2)$.  The $d_{11},d_{12},\dots,d_{1k}$ are called basic parameters while the $d_{b_i,k}$ are called functional parameters.

It remains to choose priors for the hyperparameters. \cite{2021_Rosenberger_etal} compares different commonly used prior specifications for variance priors --- inverse-gamma, uniform, and half-normal --- and found that the prior choice had little effect on point estimates.  A vague inverse-gamma prior is thus placed on $\sigma^2$, $\sigma_\mu^2$, and $\sigma_d^2$, and a vague normal is placed on the $m_\mu$ and $m_d$. Taken together, the contrast-based BNMA model with binomial outcomes for each arm is
\[
\begin{aligned}
y_{ik} \, | \, p_{ik}, n_{ik} &\sim \text{Bin}(p_{ik},n_{ik}) 
&\text{logit}(p_{ik}) &= \mu_{i} + \delta_{i,b_i,k}1_{b_i\neq k} \\
\delta_{i,b_i,k} \, &| \, \delta_{i,b_i,2},\dots,\delta_{i,b_i,k-1},
d_{b_i,2},\dots,d_{b_i,k-1}, \sigma^2 \\
&\sim \text{N}\bigg(d_{b_i,k} + \frac{1}{k-1}\sum_{j=1}^{k-1}[\delta_{i,b_i,j} - d_{b_i,j}],\frac{k}{(k-1)}\sigma^2\bigg) \\
\mu_i \, | \, m_\mu, sd_\mu &\sim N(m_\mu,\sigma_\mu^2) 
&m_\mu, m_d&\sim \text{N}(0,10000)  \\
\sigma^2, \sigma_\mu^2, \sigma_d^2 &\sim \text{IG}(1,1) 
&d_{b_i,k} &= d_{1k} - d_{1b_i} \\
d_{1k} &\sim \text{N}(m_d,\sigma_d^2)
\end{aligned}
\]

\section{Time-Varying Bayesian Network Meta-Analysis}

The studies in the dataset are indexed by $i\in\{1,2,\dots,I\}$.  The time of study $i$ is $t_i$, so that the list of possibly non-unique timepoints is $t_1,t_2,\dots,t_I$.  Treatment $k$ occurs in $I_k$ studies, and the list of studies it occurs in can be indexed by $i_k$.  The timepoints in which treatment $k$ occurs are indexed by $t_{i_k}$.  If there is a time-based design inconsistency, then $d_{b_i,k} \neq d_{1k} - d_{1b_i}$ for some studies $i$ because the basic parameters $d_{1k}$ cannot capture the time-varying nature of the treatment effect.  To remedy this, model a time-specific value of $d_{1k}$, $d_{1k}^{t_{i_k}}$, at each of the timepoints $t_{i_k}$. Then redefine the $d_{b_i,k}$: $d_{b_i,k} = d_{1k}^{t_{ik}} - d_{1b_i}^{t_{ib_i}}$. For each $k$, the $d_{1k}^{t_{i_k}}$ correspond to a latent, unobserved, potentially nonstationary time series which could exhibit any of a large number of time-varying trends.  To maintain flexibility, the $d_{1k}^{t_{ik}}$ are modelled as arising from a Gaussian Process (GP) kernel \citep{2004_Brahim_belhouari_bermak, 2006_Rasmussen_Williams}.  Let $d_{1k}^{t_{1}} \sim \text{GP}(\vect{d_{1k}},K(\cdot,\cdot))$ represent the following distribution:
\[
\begin{bmatrix}
d_{1k}^{t_{1_k}} &
d_{1k}^{t_{2_k}} &
\dots &
d_{1k}^{t_{I_{k_k}}} 
\end{bmatrix}^T
\sim 
N \bigg(
\begin{bmatrix}
d_{1k} &
d_{1k}&
\dots &
d_{1k}
\end{bmatrix}^T,
 K(\cdot,\cdot)
\bigg).
\]
Decompose the covariance kernel, $K(\cdot,\cdot)$ into three separate kernels (for more on kernel decomposition see, e.g. \cite{2020_Corani_etal}): (1) a white noise kernel, (2) a linear kernel, and (3) a Matern covariance kernel.  That is
\[
K(\cdot,\cdot) = K_{\text{W}}(\cdot,\cdot) + K_{\text{L}}(\cdot,\cdot) + K_{\text{M}}(\cdot,\cdot)
\]
The white noise kernel is
\[ 
K_{\text{W}} = \psi^2\mathbb{I}_{n_k},
\]
where $\mathbb{I}_{n_kk}$ is the $n_k\times n_k$ identity matrix.  This kernel adds white noise to the covariance terms.  The linear covariance kernel is
\[
K_{\text{L}}(i,j) = s_{bk}^2 + s_{lk}^2t_{i_k}t_{j_k},
\]
which induces linear functions in the $d_{1k}^{t_{i_k}}$.  The Matern covariance kernel, with $\nu = \frac{1}{2}$, is
\[
K_{\text{M}}(i,j) = \phi_k^2\exp(-\rho_k|t_{i_k} - t_{j_k}|)).
\]
This last kernel results in functions equivalent to the Ornstein–Ulenbeck process, the continuous time equivalent of an AR(1) model \citep{2013_Roberts_etal}. As BNMA is effective at finding the average values $d_{1k}$ (to be demonstrated below), these are taken as the mean value for the Gaussian process. Vague priors are placed on all of the hyperparameters.  Further, note that not all treatments should be modelled with time-varying effects: some treatments will not vary in time, while others will not have sufficient data to learn time-varying trends.  Let $\mathcal{T}_0$ be the set of treatments modelled as constant in time, and let $\mathcal{T}_1$ be the set of treatments modelled as varying in time.  The resulting model, termed tBNMA, is 
\[
\begin{aligned}
y_{ik} \, | \, p_{ik}, n_{ik} &\sim \text{Bin}(p_{ik},n_{ik}) 
&\text{logit}(p_{ik}) &= \mu_{i} + \delta_{i,b_i,k}1_{b_i\neq k} \\
\delta_{i,b_i,k} \, &| \, \delta_{i,b_i,2},\dots,\delta_{i,b_i,k-1},
d_{b_i,2},\dots,d_{b_i,k-1}, \sigma^2\\
&\sim \text{N}\bigg(d_{b_i,k} + \frac{1}{k-1}\sum_{j=1}^{k-1}[\delta_{i,b_i,j} - d_{b_i,j}],\frac{k}{(k-1)}\sigma^2\bigg) \\
\mu_i \, | \, m_\mu, \sigma_\mu &\sim N(m_\mu,\sigma_\mu^2) 
&d_{b_i,k} &= d_{1k}^{t_{ik}} - d_{1b_i}^{t_{ib_i}} \\
d_{1k}^{t_{ik}} \, &| \, k\in\mathcal{T}_1, d_{1k}, \psi, \phi, \rho \sim \text{GP}(d_{1k},K(\cdot,\cdot)))
&d_{1k}^{t_{ik}} \, &| \, k\in\mathcal{T}_0, d_{1k} = d_{1k}\\
K(i,j) &= K_{\text{W}}(i,j) + K_{\text{L}}(i,j) + K_{\text{M}}(i,j)
&K_{\text{W}} &= \psi^2\mathbb{I}_{nk} \\
K_{\text{L}}(i,j) &= s_{bk}^2 + s_{lk}^2t_{i_k}t_{j_k}
&K_{\text{M}}(i,j) &= \phi_k^2\exp(-\rho_k|t_{i_k} - t_{j_k}|)) \\
\psi, s_{bk},s_{lk} &\sim \text{N}_{+}(0,10000) 
& \sigma^2, \sigma_\mu^2, \sigma_d^2, \phi_k &\sim \text{IG}(1,1) \\
\rho_k &\sim \text{G}(1,1)
&d_{1k} &\sim \text{N}(m_d,\sigma_d^2) \\
m_\mu, m_d &\sim \text{N}(0,10000) 
\end{aligned}
\]
A Gibbs sampler is implemented in JAGS.  

\section{Data, Simulations, and Analysis}

MRSA-related cSSSI treatments are analyzed using the the combined data from previous studies that employed BNMA.  Using the network, treatment arms, and timepoints from these data, data is simulated with time-varying effects on one treatment.  The performances of two BNMA methods on this simulated dataset are compared to each other and to tBNMA.

\subsection{Data}

Data from nine reviews employing NMA techniques to study the efficacy of treatments for MRSA-related cSSSIs are used: \cite{2015_Thom_etal}, \cite{2016_Liu_etal}, \cite{2017_Guest_etal}, \cite{2017_McCool_etal}, \cite{2018_Li_Xu},\cite{2019_Zhang_etal}, \cite{2019_Lan_etal}, \cite{2021_Brown_etal}, and \cite{2021_Feng_etal}.  A potential concern with combining datasets from multiple studies is that they will be incompatible --- different experimental designs, for instance, may give rise to RCTs implemented on significantly different populations, violating the consistency assumption.  The reviews are all conducted according to PRISMA or Cochrane standards, so there is a measure of similarity in how they collected studies. In all of these reviews, the vast majority of studies appeared in at least one other review: this implies transitive consistency.  Given the lack of data on MRSA-related cSSSI's \citep{2021_Brown_etal}, it is better to be expansive when deciding which studies to include.  Moreover, the random effects allow the models to compensate for inconsistencies introduced by combining data from different reviews.

\begin{figure}
    \centering
    \includegraphics[width = \textwidth]{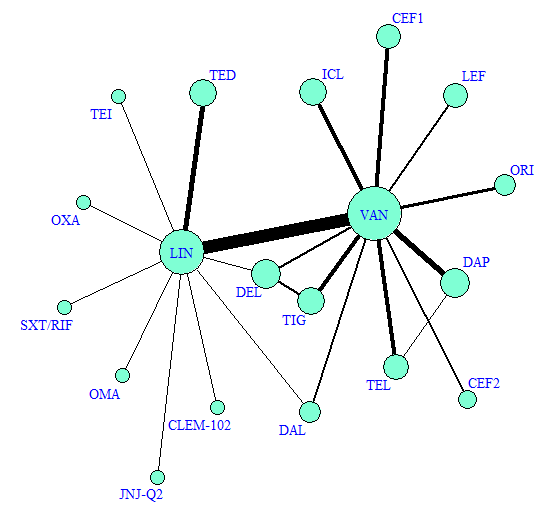}
    \caption{The network of treatments found in the agglomerated dataset.  Treatments in larger nodes appeared more often, and the thicker the line between two nodes the more often they were compared. }
    \label{fig:data_network_plot}
\end{figure}

These reviews contribute a total of $58$  studies comparing $19$ treatments from $2000$ to $2019$.  The earliest date of publication of a study is used --- if the day of publication is not available, it is imputed to be the middle of the month.  A plot of the network is provided in Figure \ref{fig:data_network_plot}.  Four studies have $3$ treatment arms; the rest have $2$.  The most prevalent treatments are vancomycin (VAN), which appears $46$ times, and linezolid (LIN), which appears $27$ times.  There are $13$ direct comparisons of the two.  Both vancomycin and linezolid have comparisons with dalbavancin (DAL) and delafloxacin (DEL), but otherwise have no common comparators and the network structure can be thought of as having two poorly connected cliques.  Vancomycin has additional comparisons with ceftaroline (CEF1),ceftobiprole (CEF2), oritavancin (ORI), daptomycin (DAP), telavancin (TEL),  tigecycline (TIG), iclaprim (ICL), and lefamulin (LEF).  Linezolid has additional comparisons with rifampicin (SXT/RIF), teicoplanin (TEI),  omadacycline (OMA), a novel fluoroquinolone  (JNJ-Q2), fusidic acid (CLEM-102), tedizolid (TED), and oxacillin-dicloxacililn (OXA).  Daptomycin and telavancin have one comparison with each other while tigecycline and delafloxin have two.  There are no other comparisons in the network.   

\subsection{Simulations}

Simulations will show the limitations of existing models in the presence of time-based designed inconsistencies, and demonstrate the ability of tBNMA to solve this problem.  The treatment comparisons, timepoints, and network associated with the combined data are used to generate the simulated data.  

Three models are compared.  The first is standard BNMA, which takes no measures beyond random effects to compensate for time-based design inconsistencies.  The second is Meta-BNMA, which runs a meta-regression on time effects by modelling the $d_{tk}^{t_{i_k}}$ as following a linear trend in time for those treatments $k$ which are allowed to vary in time.  Meta-BNMA bears similarities to models discussed by \cite{2009_Salanti_etal} and \cite{2012_White_etal}.  The third is tBNMA.  There is prior information suggesting that only two treatments present in the study design --- vancomycin \citep{2007_Daum} and fusidic acid \citep{2021_Brown_etal} --- are potentially experiencing time-varying treatment effects.  Of these, only vancomycin appears in enough studies for time-varying effects to be detectable.  Thus, the two models which account for time-based design inconsistencies, Meta-BNMA and tBNMA, will allow time-varying effects only on vancomycin.  Linezolid, the second most common treatment is used as the baseline treatment for all models.

If there are timed-based design inconsistencies, then the $d_{1k}^{t_{i_k}}$ could vary in time according to a large number of curves --- but the specific form is unknown for any given treatment $k$.  It is thus desirable to assess the performance of the three models in a number of scenarios.  Three datasets, with three different time-varying effects on the vancomycin $d_{1k}^{t_{i_k}}$, are generated.  In the first, the $d_{1k}^{t_{i_k}}$ are constant in time; in the second, the $d_{1k}^{t_{i_k}}$ are quadratic in time; in the third, the $d_{1k}^{t_{i_k}}$ are sigmoidal in time.  

\begin{figure}
    \centering
    \includegraphics[width = \textwidth]{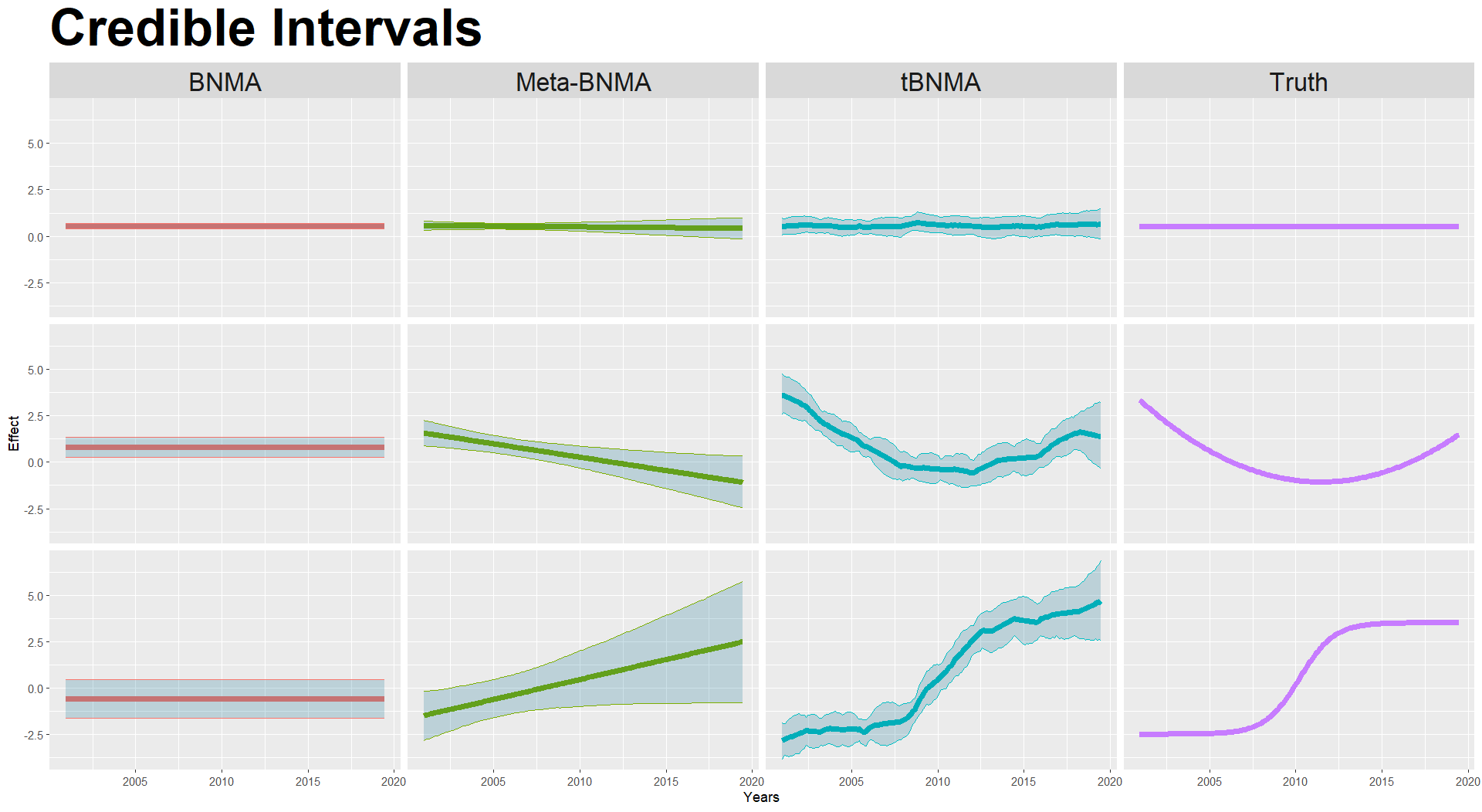}
    \caption{Posterior credible intervals for the $d_{1k}^{t_{i_k}}$ associated with vancomycin in a variety of simulated environments.}
    \label{fig:sim_vary_truth}
\end{figure}

All three models are run on all three simulated datasets.  The $95\%$ credible intervals for the posterior predictive distributions for the $d_{1k}^{t_{i_k}}$ corresponding to the relative treatment effect of vancomycin compared to linezolid over time are plotted in Figure \ref{fig:sim_vary_truth} along with the true values of the $d_{1k}^{t_{i_k}}$.  When the true curve is constant and there are no time-based design inconsistencies, all three models return approximately constant trends in time.  While all models work when there are no time-based design inconsistencies, Meta-BNMA and tBNMA have wider credible intervals than BNMA because they are more complex.  When there are time-based design inconsistencies, either quadratic or sigmoidal, there are clear differences between the models.  BNMA cannot detect time trends in the $d_{1k}^{t_{i_k}}$, though it can estimate the mean value $d_{1k}$ with considerable accuracy.  The time-based design inconsistencies result in elevated uncertainty compared to the case where the the constant function is true. Meta-BNMA detects significant time trends; however, it is limited to detecting only linear time trends and is thus unable to learn more complicated scenarios.  Moreover, it has greatly inflated credible intervals, indicating that it fits the data poorly.  In contrast, tBNMA is flexible enough to accurately recover the true time-varying effect no matter the underlying trend.

\begin{figure}
    \centering
    \includegraphics[width = \textwidth]{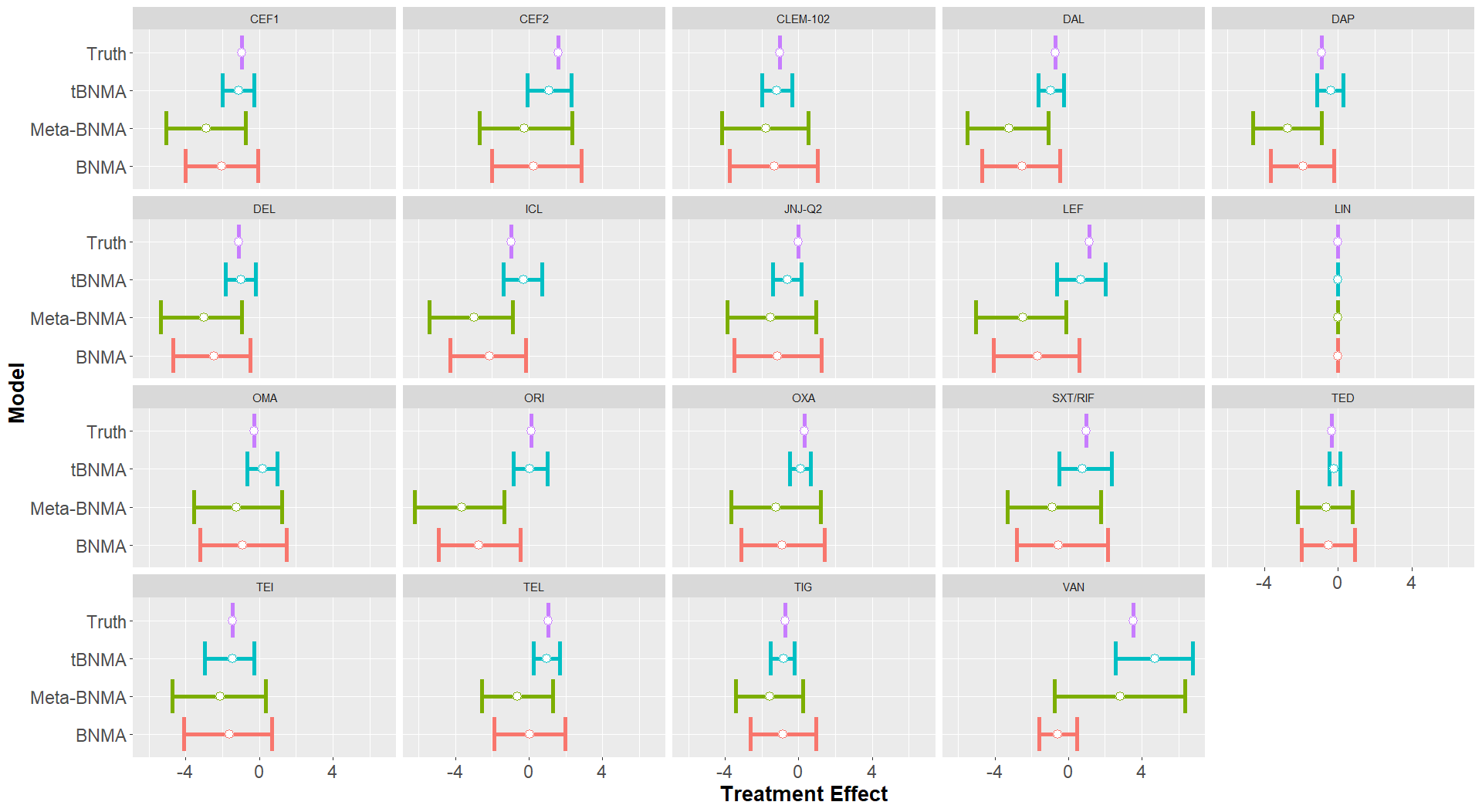}
    \caption{Point estimate and $95\%$ credible interval from the posterior predictive distribution for $d_{1k}^T$ by model when there is a sigmoidal time effect on VAN.}
    \label{fig:sim_tbnma_mean_te_ci}
\end{figure}

The better fit found by tBNMA also leads to increased predictive performance.  The quantity most of interest is $d_{1k}^T$, the relative treatment effect of treatment $k$ relative to the baseline treatment at time $T$.  As time $T$ corresponds to the end of the study period, it holds the most clinical significance.  Point estimates and $95\%$ credible intervals from the posterior predictive distributions for the $d_{1k}^T$ are found for all treatments and for all three models on the simulated sigmoidal dataset.  The results are plotted in Figure \ref{fig:sim_tbnma_mean_te_ci} along with the true values.  tBNMA consistently produces the best estimates, with the narrowest credible intervals.  Since BNMA and Meta-BNMA cannot capture the full effect of the design-based inconsistencies, they compensate by increasing the uncertainty of their predictive posterior distributions, even for treatments which do not have time-varying effects.  tBNMA thus outperforms existing methods in the presence of significant time-based design inconsistencies.

\subsection{Implementation on MRSA Data}

BNMA, Meta-BNMA, and tBNMA are run on the agglomerated dataset.  As before, linezolid is the baseline treatment for all methods.  Meta-BNMA and tBNMA allow for time-varying effects on vancomycin; all other treatment effects are fixed with respect to time. No covariates aside from time are considered because of the lack of covariate information for most of the RCTs.

\begin{figure}
    \centering
    \includegraphics[width = \textwidth]{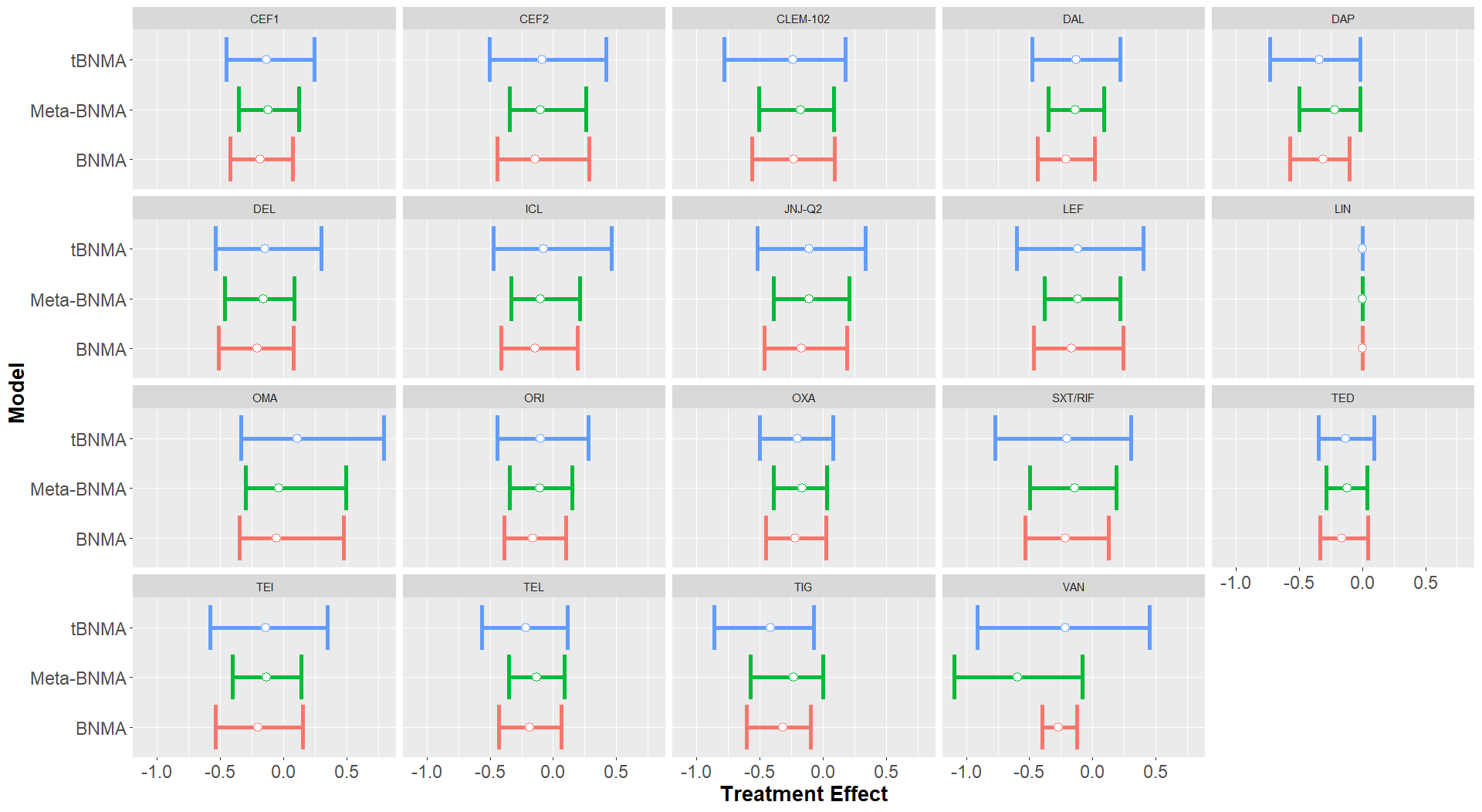}
    \caption{$95\%$ credible intervals and mean estimates for the posterior predictive distribution of $d_{1k}^T$ under BNMA, Meta-BNMA, and tBNMA on the agglomerate dataset.}
    \label{fig:data_tbnma_te_mean_cred}
\end{figure}

\begin{figure}
    \centering
    \includegraphics[width = \textwidth]{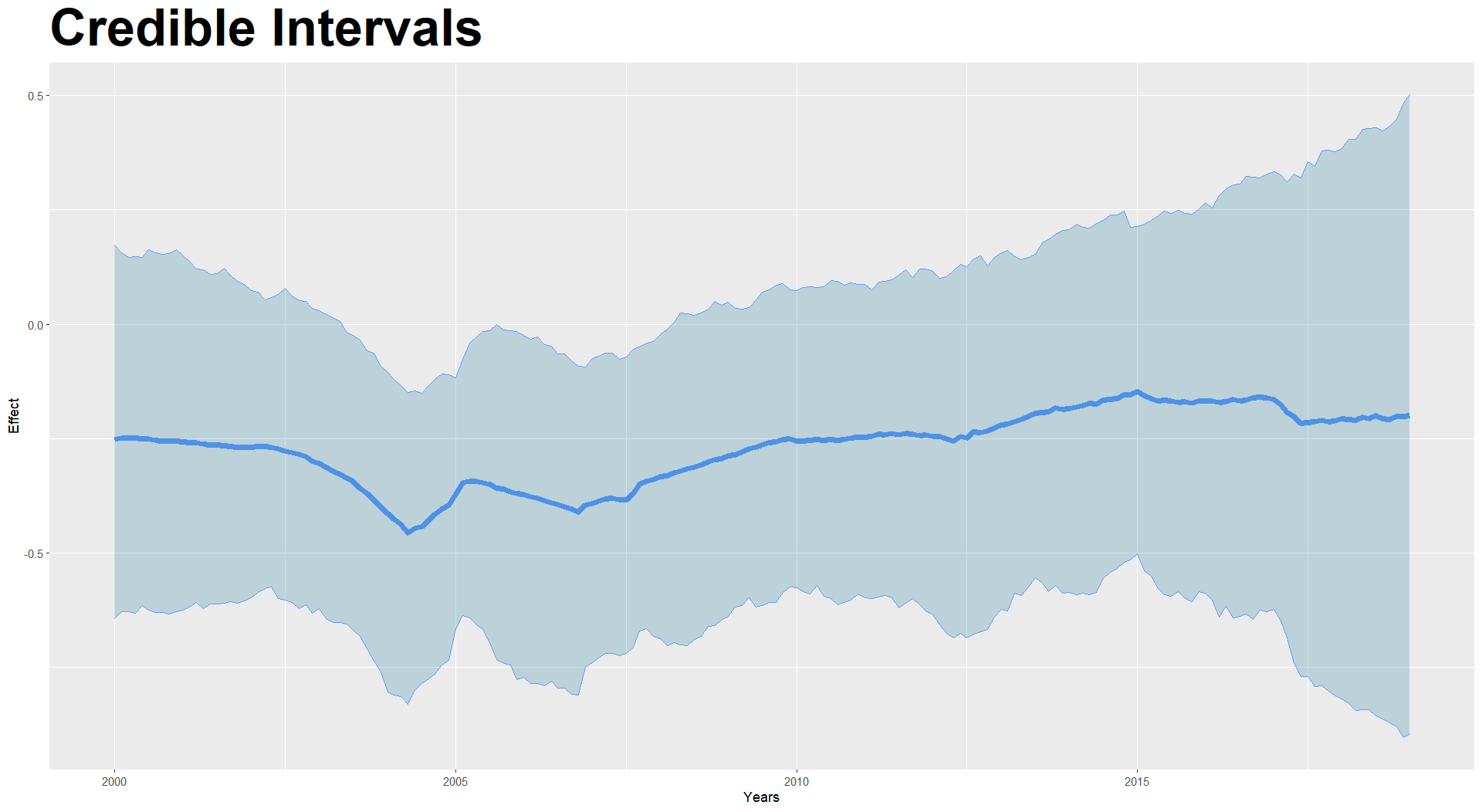}
    \caption{$95\%$ credible intervals for the posterior predictive distribution of the $d_{1k}^{t_{i_k}}$ relating the treatment effect of vancomycin to linezolid under tBNMA on the agglomerated dataset.}
    \label{fig:data_tbnma_credible}
\end{figure}

Figure \ref{fig:data_tbnma_te_mean_cred} shows the  $95\%$ credible intervals for the posterior predictive distributions of the treatment effect of each treatment relative to linezolid at the end of the time period, $d_{1k}^T$.  The three methods produce similar posterior mean estimates of the $d_{1k}^T$. Meta-BNMA and tBNMA have wider credible intervals because the time-varying effects modelled in the $d_{1k}^{t_{i_k}}$ for vancomycin induce a larger degree of uncertainty in the estimates for the other treatments.  The estimate where the models most disagree, however, is that of the treatment effect of vancomycin relative to linezolid.  That is, BNMA and Meta-BNMA find at least a $95\%$ chance that vancomycin is less effective than linezolid at treating MRSA at the end of the time period; tBNMA finds only a $75\%$ chance that this is true.  At a $95\%$ level, the models lead to different clinical inferences.  
Figure \ref{fig:data_tbnma_credible} plots the $95\%$ credible intervals for the posterior predictive distribution of the $d_{1k}^{t_{i_k}}$ relating the relative efficacy of vancomycin compared to linezolid learned by the tBNMA model.  tBNMA discovers significant non-linear trends which are not found by existing methods.  For most of the time period, vancomycin and linezolid are indistinguishable from each other at a $95\%$ level; however, vancomycin is significantly less effective from $2002$ to $2007$.  This is consistent with results from the medical literature concerning the overall prevalence of vancomycin-resistant \textit{S. Aureus}.  \cite{2004_Cosgrove_etal} and \cite{2007_Daum} reported the emergence of vancomycin-resistant \textit{S. Aureus} during this period, while \cite{2017_Klein_etal} claimed that MRSA prevalence peaked in $2008$ and \cite{2019_Diekema_etal} reported that there was no increase in vancomycin-resistant \textit{S. Aureus} from 2013 to 2016.  The most plausible explanation is that the prevalence of vancomycin-resistant \textit{S. Aureus} was rising in the mid 2000s.  Medical experts then designed and implemented a set of medical interventions designed to slow the spread of antibiotic-resistant \textit{S. Aureus} (see, e.g., \cite{2009_Liebowitz}) which tBNMA finds to be largely successful.  

Previous network-meta analyses conducted to assess various treatments for \textit{S. Aureus} have been divided on whether vancomycin is more effective than linezolid. \cite{2019_Zhang_etal}, \cite{2018_Li_Xu}, \cite{2021_Feng_etal}, and \cite{2017_McCool_etal} found that linezolid was more effective, while \cite{2015_Thom_etal} and  \cite{2017_Guest_etal} found them to be equivalent. The above results indicate that one reason for this disparity may be time-based design inconsistencies.  Standard techniques such as BNMA and Meta-BNMA found linezolid to be significantly more effective than vancomycin at the end of the time period.  However, tBNMA finds that, while linezolid used to be more effective than vancomycin at a $95\%$ level, it is not significantly more effective at the end of the time period.   Models which do not take the time-varying nature of this comparison into account may predict that there is a significant difference at the end of the time period, rather than in the middle.  

\section{Discussion}

A novel model, tBNMA, is proposed which accounts for time-based design inconsistencies in network meta-analyses of RCTs by modelling time-varying effects as a latent, unobserved, time series.  A Gaussian Process combining white noise, linear, and Matern kernels is used to model this latent series.  tBNMA is fully Bayesian and allow for posterior uncertainty quantification; posterior computation proceeds through a Gibbs sampler implemented in JAGS. tBNMA substantially outperformed existing methods in simulations in the presence of significant, non-constant, time-varying treatment effects.

Data from a collection of NMA-based review papers on MRSA-related cSSSIs is combined and analyzed using BNMA, Meta-BNMA, and tBNMA. tBNMA finds that MRSA is not more resistant to vancomycin at the end of the period  than at the beginning, but there are substantial non-linear effects.  Vancomycin resistance in MRSA was strongest between $2002$ and $2007$, in line with clinical trends, but has since declined.  The time-based nature of this disparity may account for the disagreement about whether linezolid is more effective than vancomycin in the literature. 

The time-varying methods presented in this paper could be expanded upon.  One such extension would follow \cite{2012_Jansen} or \cite{2020_Phillippo_etal} and employ a meta-regression model to ``balance" studies with covariate information to those without.  Such methods are data-intensive, however, and care would be needed to employ them simultaneously with the time-varying methods proposed in this paper.  Alternate kernels for modelling the time-varying effects could also be explored.  

\bibliographystyle{agsm}
\bibliography{sample}

\clearpage

\clearpage

\clearpage

\clearpage

\clearpage

\clearpage 

\clearpage 

\section*{Software}

Reproducible code and data for this paper is available at \url{https://github.com/PatrickLeBlanc/tBNMA}.

\end{document}